\documentclass[usenatbib]{mnras}
\usepackage[dvips]{graphicx}
\usepackage{amsmath,amsfonts,amssymb}
\usepackage{color}
\usepackage{bm}
\usepackage{amssymb}
\usepackage{float}
\usepackage{hyperref}
\hypersetup{colorlinks=true,
           citecolor=blue,
           linkcolor=blue}
\definecolor{orange}{rgb}{1,0.5,0}
\def\icarus{Icarus}
\def\aaps{A\&AS}

\def\aj{Astron.\ J. }
\def\apj{Astrophys.\ J. }
\def\apjl{Astrophys.\ J. }
\def\apjs{Astrophys.\ J.\ (Supp.) }
\def\aap{Astron.\ Astrophys. }

\def\mnras{Mon.\ Not.\ R.\ Astron.\ Soc. }

\def\nat{Nature }

\def\sovmath{Sov.\ Math.\ Dokl.}

\newcommand{\be}{\begin{equation}}
\newcommand{\ee}{\end{equation}}
\newcommand{\bea}{\begin{eqnarray}}
\newcommand{\eea}{\end{eqnarray}}

\newcommand{\st}[2]{\ensuremath{#1_{\textrm{#2}}}}

\begin{document}

\title[Diffusion in Gliese-876]{Chaotic Diffusion in the Gliese-876 Planetary System}
\author[Mart\'i, Cincotta \& Beaug\'e]{\large{J. G. Mart\'i$^{1,2}$, P.M. Cincotta$^{1}$ C. Beaug\'e$^{2}$} \\
$^{1}$Grupo de Caos en Sistemas Hamiltonianos,
Facultad de Ciencias Astron\'omicas y
Geof\'{\i}sicas, Universidad Nacional de La Plata and\\
Instituto de Astrof\'{\i}sica de La Plata (CONICET-UNLP),
La Plata, Argentina\\$^{2}$Instituto de Astronom\'ia Te\'orica y Experimental (IATE), Observatorio Astron\'omico, Universidad Nacional de C\'ordoba, Argentina
}

\date{}
\maketitle

\begin{abstract}
Chaotic diffusion is supposed to be responsible for orbital instabilities in planetary systems after the dissipation of the protoplanetary disk, and a natural consequence of irregular motion. In this paper we show that resonant multi-planetary systems, despite being highly chaotic, not necessarily exhibit significant diffusion in phase space, and may still survive virtually unchanged over timescales comparable to their age.Using the GJ-876 system as an example, we analyze the chaotic diffusion of the outermost (and less massive) planet. We construct a set of stability maps in the surrounding regions of the Laplace resonance. We numerically integrate ensembles of close initial conditions, compute Poincar\'e maps and estimate the chaotic diffusion present in this system. Our results show that, the Laplace resonance contains two different regions: an inner domain characterized by low chaoticity and slow diffusion, and an outer one displaying larger values of dynamical indicators. In the outer resonant domain, the stochastic borders of the Laplace resonance seem to prevent the complete destruction of the system. We characterize the diffusion for small ensembles along the parameters of the outermost planet. Finally, we perform a stability analysis of the inherent chaotic, albeit stable Laplace resonance, by linking the behavior of the resonant variables of the configurations to the different sub-structures inside the three-body resonance.
\end{abstract}


\section{Introduction}\label{sec1}

Planetary systems constitute a paradigm of classical N-body problems. It has long been known that a general N-body system with $N \ge 3$ is not integrable. \citet{arnold_1963} showed that a typical near-integrable Hamiltonian system (HS) with more than 2 degrees of freedom is topologically unstable, even for a negligible value of the perturbation. Thus, given a sufficiently long period of time, the actions in the phase-space could diffuse from their initial values and lead to orbital instabilities. However, estimates for the instability time-scales are given just for extremely small perturbations \citep{neckhoroshev_1977, chirikov_1979, cincotta_2014}, being exponentially large. General estimates of diffusion time-scales for low-to-moderate perturbations are still lacking.   

In planetary systems, the diffusion timescale may be a strong function of the initial conditions, particularly in the vicinity of mean-motion resonances. Thus, how long a system can last until completely destroyed is an unsolved problem with great astronomical interest \citep{laskar_1989}.
In Hamiltonian systems, orbital instabilities (and, consequently, strong chaotic diffusion) are generated by the overlap of resonances \citep{wisdom_1980}, and planetary dynamics are no exception. Although 
many works in recent times have tried to establish a relationship between chaos and instability \citep{marchal_saari_1975, marchal_bozis_1982, chambers_1996, smit_lissauer_2009, giuppone_morais_correia_2013, deck_etal_2013, ramos_etal_2015}, no general results have been so far obtained, particularly for the case $N > 2$.

As the number of detected exoplanets increased, so did their orbital diversity. Short period with nearly circular orbit planets are supposed to have undergone large scale orbital migration from beyond the snow line, where giant planets are known to be formed. Many of these short period planets are so close to their parent star that tidal dissipation would have likely circularized their orbits \citep{marti_beauge_2015}. Thus, current orbital parameters of such bodies do not provide a good indicator of their dynamical history. 

On the other hand, planets in eccentric orbits are generally believed to have formed on nearly circular orbits and later evolved to their presently observed large eccentricities. Among the proposed mechanisms for producing large eccentricities are a passing binary star \citep{laughlin_adams_1998}, secular perturbations due to a distant stellar or planetary companion \citep{ford_etal_2000} and strong planet-planet scattering events \citep{rasio_ford_1996, weidenshilling_marzari_1996, juric_tremaine_2008, beauge_nesvorny_2012}. 

Multi-resonant configurations are supposed to be a natural outcome of disk-driven planetary migration \citep{masset_snellgrove_2001, morbidelli_etal_2007ii,hands_dehnen_2014}, and their orbital features are not believed to have been affected by planetary instabilities such as planet-planet scattering or Lidov-Kozai resonance. Thus, their configurations should be more representative of the end product of the formation process, and thus indicative of the stability of ``dynamically quiet'' systems. 

Among the population of resonant and near-resonant systems \citep{rivera_etal_2010, fabrycky_etal_2012, wang_etal_2012, marti_giuppone_beauge_2013, rowe_etal_2014}, a large number has been discovered by the \emph{Kepler} mission. However, some of these are still awaiting confirmation, and several key orbital parameters (including their masses) are not known. Thus, in order to perform a detailed dynamical analysis of resonant systems, it seems preferable to turn to those radial velocity detections in which the inclination of the orbital plane has been (at least qualitatively) estimated. One of the best choices is GJ-876, and will be used as our main target in our analysis of diffusion in extrasolar multi-resonant planetary systems.

The GJ-876 system contains, up to date, four confirmed planets orbiting an M-type central star ($M_{\bigstar}$ from 0.32 to 0.334 $M_{\odot}$ depending on author). The inner planet (GJ-876 d) is  
very small, located very close to the star, and dynamically detached from the rest of the system. The three other planets are known to be in the vicinity of a Laplace-type resonance, and have been the subject of several investigations (e.g. \citet{rivera_etal_2010, baluev_2011, marti_giuppone_beauge_2013, batygin_holman_2015}). 

A detailed dynamical analysis of this system was presented in \citet{marti_giuppone_beauge_2013}, where it was shown that the multi-resonant configuration displayed by GJ-876 is chaotic, albeit long-term stable. In that paper we presented a series of dynamical maps and found stability limits on the  mass ratio of the outer planets as well as precise boundaries on the mutual inclination of the system, inferring that the most likely dynamically relaxed configuration is the co-planar case. Most important, once acknowledged that the system is actually multi-resonant, we retrieved specific values for the angular parameters of the planets to ensure a better representation for the plane of initial conditions. In this way we were able to fix initial angular variables in order to define a representative plane obtained via dynamical considerations, where the Laplace resonance can easily be identified.

In this work we aim to give a qualitative picture of the different chaotic processes (regimes) that can be explored by the three-body resonant configuration depicted by the paradigmatic GJ-876 system. Through this, we want to quantify the variation of the actions of the system, associated with fundamental orbital parameters of the planets, by means of a realistic numerical computation of the diffusion coefficients.

\section{Chaotic diffusion}
\label{chaos0}

\subsection{Summary of resonant perturbation theory}
\label{chaos1}

In order to sketch the geometry of resonant dynamics in action space, following \citet{chirikov_1979} and \citet{cincotta_2002}, let $\bm{I}$ denote the $N$-dimensional action vector and $\bm{\theta}$ its conjugate canonical $N$-dimensional angle, and $H_0(\bm{I})$ the unperturbed integrable non-linear Hamiltonian. Then the frequency vector $\bm{\omega}(\bm{I})=\nabla_{\bm{I}}H_0$ is always normal to the unperturbed energy surface $H_0(\bm{I})=h$. The resonance condition  $\bm{k}\cdot\bm{\omega}(\bm{I^r})=0$,  where $\bm{k}$ is a non-zero 
$N$-dimensional vector of integers and $\bm{I^r}$ the resonant action, leads to the resonance surface $\Sigma_{\bm{k}}$. Thus on any resonant torus, the resonant vector $\bm{k}$, is tangent to the energy surface.

Any perturbation to $H_0(\bm{I})$, $\varepsilon V(\bm{I},\bm{\theta})$, where $\varepsilon\ll 1$ and $V$ is an analytic function introduces variations in the unperturbed actions or global integrals. The latter can be Fourier expanded in the angular variables with coefficients that depend on the actions as:
$$\varepsilon V=\varepsilon\sum_{\bm{k}\ne 0}V_{\bm{k}}(\bm{I})\exp{(i\bm{k}\cdot\bm{\theta})}.$$

In the single resonance formulation, for sufficiently small $\varepsilon$ and initial conditions such that the system is close to the resonance $\bm{m}\cdot\bm{\omega}(\bm{I^r})=0$, retaining only
the largest (real) term corresponding to the resonant phase, $\bm{m}\cdot\bm{\theta}$, averaging out all the remaining ones we get for $|\bm{I}-\bm{I^r}|\lesssim 2\sqrt{\varepsilon}$ the local Hamiltonian 
\begin{equation}
H(\bm{I},\bm{\theta})=H_0(\bm{I})+\varepsilon V_{\bm{m}}(\bm{I})\cos(\bm{m}\cdot\bm{\theta}),
\label{eq1}
\end{equation}
and thus
\begin{equation}
\dot{\bm{I}}=-\frac{\partial H}{\partial\bm{\theta}}=\varepsilon \bm{m}V_{\bm{m}}(\bm{I})
\sin(\bm{m}\cdot\bm{\theta}).  
\label{eq2}
\end{equation}
The above relation shows that the variation of $\bm{I}$ has the direction of the resonant vector $\bm{m}$, tangent to the energy surface.

Since the motion is one-dimensional, it is possible to introduce a canonical local change of coordinates (or local change of basis) around
$\bm{I^r}$: 
$(\bm{I},\bm{\theta})\to(\bm{J},\bm{\psi})$ such that
$\psi_1=\bm{m}\cdot\bm{\theta}$, 
and $\bm{I}=\bm{I^r}+\bm{m}J_1$, where 
$J_1\lesssim |\bm{I}-\bm{I^r}|\sim\mathcal {O}(\sqrt{\varepsilon})$. Since the resonant Hamiltonian is cyclic in $\psi_2,\cdots, \psi_N$, we can neglect $J_2;\cdots,J_N$ and then 
keeping terms up to $J_1^2$,
it takes the well-known pendulum form
\begin{equation}
H_r(J_1,\psi_1)=\frac{J_1^2}{2M}+V_{\bm{m}}(\bm{I^r})\cos\psi_1
\label{eqHr}
\end{equation}
where 
$$M^{-1}=\sum_{i,j}
m_{i}\left(\frac{\partial\omega_i}
{\partial I_j}\right)_{\bm{I^r}}\!\!m_{j},$$
is the inverse of the non-linear mass, assumed to be different from zero. All the $N-1$ actions $J_2,\cdots, J_N$ are local integrals of the motion whose numerical values should be zero for $\bm{I^r}$ to be an allowed value for the perturbed motion. While $J_1$ is the action component in the direction of $\bm{m}$, $J_2$ could be taken normal to the energy surface (in the direction of
$\bm{\omega}(\bm{I^r})\equiv\bm{\omega^r}$) and thus motion in $J_2$ could be ignored. The remaining $N-2$ components, $J_3,\cdots, J_N$ belong to the $N-2$ dimensional manifold, the diffusion manifold,  defined
by the intersection of the energy and resonance surfaces.

Now, let us discuss a crucial difference between HS with $N\le 2$ and $N>2$ degrees of freedom. 

In low-dimensional non-degenerated HS, for instance $N=2$, the unperturbed 
energy surface $H_0(I_1,I_2)=h$ is 1-dimensional, just a curve. The resonance surface
$m_1\omega_1(I_1,I_2)+m_2\omega_2(I_1,I_2)=0$ is also 1-dimensional. Therefore the intersection
of both, energy and resonance surfaces is a single point, $(I_1^r, I_2^r)$, a unique torus, the resonant torus. Then the motion takes place along  the resonant vector $\bm{m}$, tangent to the energy surface. Thus due to the dimensionality of the energy surface and the invariant tori, any transition from one torus to another is only possible through all the intermediate tori between them. Thus the motion under a single resonant perturbation is tangent to the energy surface (curve) and \emph{transverse} to the resonance surface (curve). Since the dense set of resonance surfaces do not intersect each other over the energy surface, large chaos and possibly diffusion is only possible if the perturbation is large enough such that the overlap of nearby resonances takes place. For very small perturbations, chaos is just confined to the thin chaotic layers around the unperturbed separatrix of any resonance and thus the motion is mostly stable.   

For $N$--dimensional HS with $N\ge 3$,  the intersection of energy and resonance surfaces has dimension $N-2\ge 1$. Now it is clear that the set of all resonance surfaces intersect over the whole energy surface, leading to the so-called \emph{Arnol'd web}. Focusing again on an isolated resonance, since the motion is confined to the energy surface and has the direction of $\bm{m}$ ($J_1$), there are $N-2$ additional directions where motion could proceed when considering the effects of the perturbing terms in the Fourier expansion of $\varepsilon V$ (besides the resonant one). 
For instance, when $N=3$ the remaining direction could be taken \emph{along} the direction of the intersection of the energy and resonance surface. This additional direction for the motion corresponds to the third component of the local action $\bm{J}$, $J_3$. For $\varepsilon$ small enough and initial conditions such that $\bm{I}\approx\bm{I^r}$, retaining all (or at least the two largest) perturbing terms in the Fourier expansion (besides the resonant one), a slightly perturbed pendulum model is expected, with a thin chaotic layer instead of a smooth separatrix. And moreover, motion in $J_3$--\emph{along the resonance} could also take place. 
 
It has been conjectured that any orbit lying in this thin chaotic layer might visit the 
the whole Arnol'd web \citep{arnold_1964}. Arnol'd showed the existence of motion along the chaotic layer of a given resonance in a rigorous way, for a rather simple near-integrable 3D Hamiltonian.
He proved that for a  small enough perturbation it is possible to find a trajectory in the vicinity of the separatrix of a given resonance that connects two points separated by a finite distance, i.e. independent of the size of the perturbation but on a very long timescale. Arnol'd's proof rests on the existence of a chain of tori along the center of this resonance that provide a path for the orbit. If these tori are very close to each other, this orbit could transit over the chain. Since every torus in the chain is labeled by an action value, a large but finite variation of this action could take place. This mechanism, which permits motion along the resonance chaotic layer, is known as the \emph{Arnol'd Mechanism}, while the term \emph{Arnol'd diffusion} generally refers  to a possibly global phase-space instability \citep{giorgilli_1990, lochak_1999, cincotta_2002}, that is any (chaotic) orbit might visit the full Arnol'd web in a finite time.
However the problem of how to extend Arnol'd mechanism to a generic Hamiltonian remains unsolved. One of the main difficulties is related to the construction of such a chain of tori. 

Regardless of this severe limitation to understand Arnol'd diffusion as a global instability, it was largely assumed that Arnol'd diffusion does occur, and it is responsible for the chaotic mixing in relatively large regions of phase space. Nevertheless, in spite of the mathematical difficulties in dealing with this conjecture as a global instability, a local formulation shows that 
exponentially large times are necessary in order to observe any appreciable variation of the unperturbed integrals. This suggests that Arnol'd diffusion should be irrelevant in actual systems.

On the other hand, those systems exhibiting a divided phase space, where the chaotic component is relevant (and not only confined to the chaotic layers), the timescale for any diffusion (not Arnol'd diffusion) would be much shorter but still very long \citep[see for instance][]{chirikov_1997, giordano_cincotta_2004}, like a power law on the perturbation parameter. In the limit of completely random motion, this time-scale -- the inverse of the diffusion coefficient -- should go as $\sim\varepsilon^{-2}$. When resonance overlap takes place, any description such as the Arnol'd Mechanism is no longer possible since the connected resonance domains become almost completely chaotic, and the required chain of tori does not exist. So, we should use numerical experiments to quantify any diffusion.

\subsection{Diffusion}
\label{chaos2}

In this section we discuss the so-called chaotic mixing. In terms of the planetary orbits, roughly speaking, chaotic mixing means that trajectories starting in a very small neighborhood of a given point in phase space, will loose their memory about initial conditions and, for large enough times, all these trajectories appear uncorrelated. This expected ``random'' behavior could be described as a diffusion process in action space. In the limit of a Brownian type motion, the variance of any action grows linearly with time and thus, a local diffusion coefficient could be defined as the constant rate at which the variance changes with time.

However, in any realistic HS the dynamical behavior is rather far from a completely random motion. Thus, in order to characterize and quantify diffusion we should proceed with numerical experiments. Assume we are dealing with a 3D HS, which can be described in the following action-angle variables: $(\bm{I},\bm{\vartheta})$. Perform a dynamical map with any chaos indicator over a large set of initial conditions, for instance taking a grid on the $(I_1, I_2)$ plane, and keeping fixed the values of $\vartheta_i=\vartheta_i^0, i=1,2,3,$ and $I_3=I_3^0$. Any chaos indicator will provide information about the local exponential divergence around any given point of the full phase space, in this case represented by the plane where we let the initial values of the actions vary, $(I_1, I_2)$. 

With this dynamical information at hand, let us consider an ensemble of $n_p$ initial conditions in a small neighborhood of size $\sigma$ around a given point $(I_1^* I_2^*)$ on the plane $(I_1^0,I_2^0)$ with the very same values for the remaining variables, $\vartheta_i=\vartheta_i^0, I_3=I_3^0$ and where the indicator reveals an unstable, chaotic behavior. We integrate the equations of motion for all the $n_p$ points in the ensemble. The space and time distribution of all the points in $\sigma$ would give us information about the relevance of diffusion for that point. Moreover we could compute the time evolution of the space variance of the two action components distributions.

As it was already shown in \citep{cincotta_2014}, the above mentioned variance computation should be done after performing a sequence of canonical transformations to a ``good'' set of variables. Indeed, in that work it was shown that using the original set of actions, particularly when the perturbation is small, stable oscillations could hide the slow secular growth of the variance with time and thus the local diffusion coefficient would be largely underestimated. However this normal form computation to get the appropriate set of variables is not easy to be done in general, and since we will not deal with very small perturbations, we adopt an alternative way \citep{guzzo_etal_2006,lega_etal_2003}, to reduce somewhat the effect of oscillations in the drift. The above mentioned procedure to measure the diffusion in the action plane means considering a section of phase-space such that all initial conditions starting in $\sigma$ should satisfy at a given time $t$: $$|\vartheta_1(t)-\vartheta_1^0|+|\vartheta_2(t)-\vartheta_2^0|+|\vartheta_3(t)-\vartheta_3^0|< \delta_1,\quad
|I_3(t)-I_3^0|<\delta_2,$$
with $\delta_1,\delta_2\ll 1.$ This procedure, though computational expensive, will effectively
reduce the presence of fast periodic oscillations in the time evolution of the action variances.

\section{Hamiltonian Model for a Three-Body Resonance} \label{model}

Let us consider a system of three planets (masses $m_1$, $m_2$ and $m_3$) orbiting a star $m_0$ under their mutual gravitational forces. The index is chosen such that the initial semi-major axes satisfy the condition $a_1 < a_2 < a_3$.

A canonical set of variables introduced by Poincar\'e allows us to write the Hamiltonian for the four-body problem. Following \citet{laskar_robutel_1995}, let $\mathbf{r}_{i}$ be the astrocentric positions of the planets, and $\mathbf{p}_{i}$ be the barycentric momentum vectors. The pairs $(\mathbf{r}_{i},\mathbf{p}_{i})$ form a canonical set of variables with the Hamiltonian given by:
\begin{equation}
 H = H_{0} + \st{H}{dir} + \st{H}{kin}.
\label{hamil}
\end{equation}
Here $H_{0}$ is the keplerian part while the perturbations are given by the two remaining terms.  $H_{\textrm{dir}}$ is the direct part, and $H_{\textrm{kin}}$ is the kinetic part of the Hamiltonian, each expressed in terms of the canonical $(\mathbf{p}_{i},\mathbf{r}_{i})$ variables as
\begin{equation}
\begin{split}
 H_{0}         =& -\sum_{i=1}^{3}\left(\frac{p_{i}^{2}}{2\beta_{i}} - \frac{m_{0}m_{i}}{r_{i}}\right)\\
 \st{H}{dir}  =& -{\cal G} \sum_{i,j=1 \; i \ne j}^{3}\frac{m_{i}m_{j}}{\Delta_{ij}}\\
 \st{H}{kin}  =& \sum_{i,j=1 \; i \ne j}^{3}\frac{\mathbf{p}_{i}\cdot\mathbf{p}_{j}}{m_{0}},
\label{hamiltonian}
\end{split}
\end{equation}
where $\beta_{i} = m_{0}m_{i}/(m_{0} + m_{i})$, $\Delta_{ij} = |\vec{r_{i}} - \vec{r_{j}}|$, and ${\cal G}$ denotes the gravitational constant. The first term of Eq. \eqref{hamil} defines the Keplerian motion of each planet around the star, while \st{H}{dir} and \st{H}{kin} represent the mutual interactions among the planets. The barycentric momentum $\mathbf{p}_{i}$ in the four-body problem are defined as
\begin{equation}
 \mathbf{p}_{i} = \frac{m_{i}}{m_{T}}\left[  \dot{\mathbf{r}}_{i} \sum_{j \ne i} m_{j} - \sum_{j \ne i}m_{j} \dot{\mathbf{r}}_{j}\right],
\end{equation}
where $\dot{\mathbf{r}}_{i}$ are the derivatives of the astrocentric positions and $m_{T}=\Sigma_{i=0}^{3}m_i$. Since we are assuming co-planar orbits, our system contains a total of six degrees of freedom.

Performing a canonical transformation to the modified Delaunay variables, which for the planar case are given by
\begin{equation}
 \begin{split}
  & L_{j} = \beta_{j}\sqrt{\mu_{j}a_{j}} \\
  & S_{j} = L_{j}(1-\sqrt{1 - e_{j}^{2}})
 \end{split}
\label{2}
\end{equation}
with $\mu_{j}\!=\!{\cal G}(m_{0}+m_{j})$, the Keplerian part of the Hamiltonian is simply given by the expression:
\begin{equation}
 H_{0} = -\sum_{i=1}^{3}\frac{\mu_{i}^{2}\beta_{i}^{3}}{2L_{i}^{2}}.
\label{keplerian}
\end{equation}

In the vicinity of a Laplace-type resonance, we introduce new angular variables in terms of the primary resonant angles for each of the single resonances:
\begin{equation}
 \begin{split}
  \sigma_{1} &= 2\lambda_{2} - \lambda_{1} - \varpi_{1}\\
  \sigma_{2} &= 2\lambda_{3} - \lambda_{2} - \varpi_{2}\\
  \Delta\varpi_{1} &= \varpi_{2} - \varpi_{1}\\
  \Delta\varpi_{2} &= \varpi_{3} - \varpi_{2}.\\
 \end{split}
\label{4}
\end{equation}
The resonant angle of the Laplace resonance may be written in terms of the mean longitudes as:
\begin{equation}
\phi_{lap} = \lambda_{1} - 3\lambda_{2} + 2\lambda_{3} .
\end{equation}
After an averaging process with respect to the short-period terms, the resulting resonant Hamiltonian reduces to a system of four degrees-of-freedom.

\section{Dynamical Maps}

\subsection{Numerical Setup}
\label{sec4.1}
For all our numerical runs we used an N-body code based on a Bulirsh-Stoer integrator with a variable step-size in order to control the relative error ($E_{r}$) in each time-step. This value was taken equal to $E_{r} = 10^{-12}$.

We constructed a series of dynamical maps using a rectangular grid of initial conditions in the representative plane $(a_{3},e_{3})$. All other variables, as well as the planetary masses, were taken from Table \ref{table2}, which correspond to values of the angles that lead to minimum excursions in the eccentricities (see \citet{marti_giuppone_beauge_2013} for details). 

\begin{table}
\centering
\begin{tabular}{ l c c c }
\\[1ex] 
\hline\hline \\[-1.3ex]
\multicolumn{4}{c}{Orbital Parameters for the GJ-876 system} \\ [1ex]
\hline\\
{\bf Parameter} & {\bf Planet c} & {\bf Planet b} & {\bf Planet e} \\
\hline \\
   $P$ (days)                 & $30.0881$  & $61.1166$  & $124.26$ \\ 
   $m \, (\textrm{M}_{jup})$  & $0.7142$   & $2.2756$   & $0.0459$ \\
   $a$ (AU)                   & $0.129590$ & $0.208317$ & $0.3343$ \\ 
   $e$                        & $0.25591$  & $0.0324$   & $0.055$  \\ 
   $\varpi \, (^{\circ})$     & $0.0$      & $0.0$      & $180.0$  \\
   $M \, (^{\circ})$          & $240.0$    & $120.0$    & $60.0$   \\
\hline
\end{tabular}\label{table2}
\\[1ex]
\caption{
Masses and orbital elements for the three planets of GJ-876 involved in the Laplace resonance. The values of the angular variables ($\varpi$ and $M$) were chosen to minimize the variations of the orbital elements over time, and lead to small-amplitude librations of the resonant angles. The $(a_3,e_3)$ values correspond to those obtained by the four-planet co-planar fit in \citet{rivera_etal_2010}.}
\end{table}

\begin{figure}
\centering
\mbox{\includegraphics[width=8.0cm]{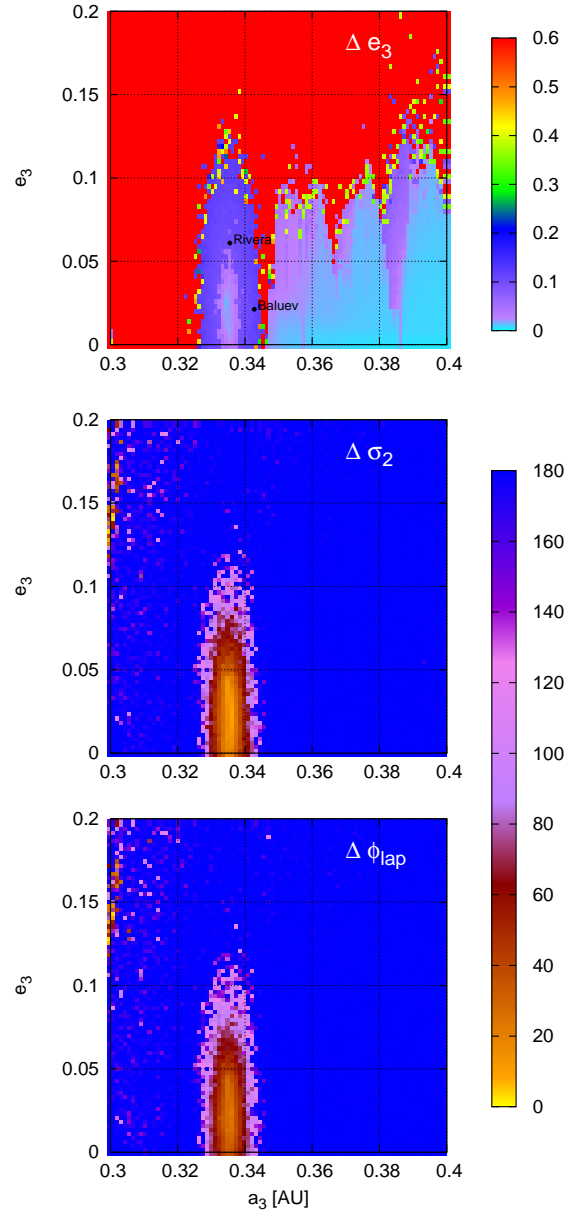}}
\caption{Top frame shows a $\Delta e_3$ dynamical map in the vicinity of the 2/1 MMR between $m_3$ and $m_2$ (corresponding to $a_3 \simeq 0.335$ AU). The middle plot shows the amplitude of libration of the primary resonant angle $\sigma_2$ of the two-body resonant, while the bottom graph shows the amplitude of libration of the Laplace resonance.}
\label{fig1}
\end{figure}

The top frame of Figure \ref{fig1} reproduces the structure of the phase-space in the $(a_3,e_3)$ representative plane in the vicinity of the 2/1 mean-motion resonance (MMR) between $m_3$ and $m_2$. Black symbols correspond to the orbital fits of \citep{rivera_etal_2010, baluev_2011}, each numerically integrated in order to intersect the representative plane. The dynamical map was constructed with a $82 \times 82$ grid, and each initial condition was integrated for $5 \times 10^4$ years. The plot shows the value of $\Delta e_3$ obtained during this time-span, with a color code in the range of $0.0<\Delta e_{3}<0.6$. The region associated to the 2/1 commensurability is clearly seen around $a_3 \simeq 0.335$ AU, while other resonances are also detected for larger semi-major axis. This plot is analogous to Figure 7 of \citet{marti_giuppone_beauge_2013}.

Initial conditions identified with red correspond to unstable orbits that lead to a disruption of the system within the integration time-span. Stable orbits in the vicinity of the 2/1 MMR define a horse-shoe type region with eccentricity reaching up to $e_3 \simeq 0.1$. Close to the stability boundary, the values of $\Delta e_3$ are relatively large (of the order of $0.2$). We also identified, deep inside the resonance domain, a small region characterized by very low eccentricity variations.

The two lower graphs show the semi-amplitude of libration of $\sigma_2$ (middle plot) and of the Laplace angle $\phi_{lap}$ (lower plot). Both show very similar behavior, indicating that practically all initial conditions within the 2/1 MMR also correspond to motion within the Laplace multi-planet resonance. 

Moreover, the region within the Laplace resonance with $\Delta e_3 \sim 0$ corresponds to small-amplitude librations of its critical argument, as expected.

\subsection{Structure of the Laplace Resonance}
\label{sec4.2}
In order to realistically assess the chaotic diffusion of this system, we must first define the basic configurations with which to compare the time-evolved parameters.

The best fit for the 3-body GJ-876 system is, according to a variety of works \citep{rivera_etal_2010, baluev_2011}, a chaotic condition; however, it has also been established that the configuration is stable and locked in a resonant state for extremely long timescales. In \citet{marti_giuppone_beauge_2013} we presented a thorough exploration of the parameter space, yielding several dynamical constrains. 

For instance, we concluded that both dynamical tests and stability considerations point towards a co-planar configuration. We also showed that finite masses are necessary in order to guarantee stability, and estimated upper bounds for the mass ratio. Here we expand on those results and discuss in more details the evolution of both regular and chaotic orbits with a higher resolution.

\begin{figure}  
\centering
\includegraphics[width=8.0cm]{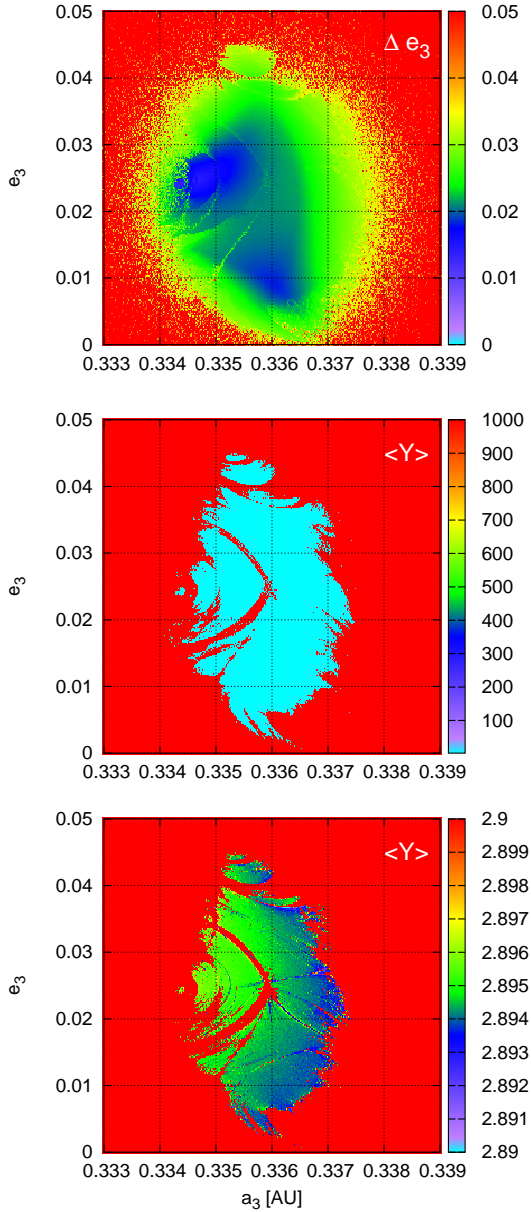}
\caption{Dynamical maps in the representative plane $(a_3,e_3)$ in the vicinity of the Laplace resonance. The color code in the top frame corresponds to $\Delta e_3$ while the two remaining graphs plot values of the MEGNO indicator $\langle Y \rangle$.}
\label{fig2}
\end{figure}

Figure \ref{fig2} presents new dynamical maps for the central region of the Laplace resonance, corresponding to low-amplitude librations of $\phi_{lap}$. Since we are interested in a detailed analysis of the resonance structure, we increased the resolution to a $300 \times 250$ grid of initial conditions in the $(a_3,e_3)$ plane. The total integration time was also increased to $10^{5}$ years. The values of $\Delta e_{3}$ for each initial condition are shown in the top panel, with a color code in the range of $0.0<\Delta e_{3}<0.06$. 

It is important to recall that $\Delta e_{3}$ is not a chaotic indicator (e.g.\citet{ramos_etal_2015}), although it constitutes an important tool with which to map changes in the structure of the phase space, such as those stemming from separatrix crossings. The MEGNO indicator \citep{cincotta_2000,cincotta_giordano_simo_2003}, on the other hand, is a robust and efficient chaos indicator. 

The middle panel of Figure \ref{fig2} shows the same map although this time the colors correspond to the MEGNO values $\langle Y \rangle$, where $2$ the lowest value, indicates regular motion. We note a very sharp transition between moderate values close to $2$ deep within the libration domain, and highly chaotic motion with $\langle Y \rangle \ge 1000$. The low-Megno region is located in the core of the resonant domain and corresponds to small-amplitude librations of the Laplace angle, as discussed in Figure \ref{fig1}.

Although both indicators do not show exactly the same results, they share some qualitative features. In both cases, the phase-space appears separated into two distinct regions: a moderately regular ($\langle Y \rangle < 3$) domain surrounded by a significantly more chaotic region identified with $\langle Y \rangle > 10$. Hereafter we will refer to each as the {\it inner} and {\it outer} resonant domains, respectively. 

The lower frame presents, once again, a MEGNO color map, only this time limited to values found in the inner core of the resonance. We can now see a number of dynamical structures deep within this commensurability. Although similar structures may also be seen in the $\Delta e_3$ map, these are not so clearly defined. A second interesting result of the MEGNO map is that all initial conditions appear chaotic (reaching a minimum value of $\langle Y \rangle \simeq 2.89$), even for very low amplitudes of libration. Moreover this figure clearly
shows the very signatures of high order resonances within this domain as narrow channels or simply as
smooth curves (see below).

This general chaoticity is not unexpected. Indeed, \citet{nesvorny_morbidelli_1999} considered the full three-body-resonance as a configuration in the SS system (asteroid, Jupiter and Saturn), in which the time derivative of a generic resonant angle $\sigma$ satisfies: 
\begin{equation}
\dot{\sigma} = j_1\dot{\lambda}_1 + j_2\dot{\lambda}_2 + j_3\dot{\lambda}_3 + l_1\dot{\varpi}_1 + l_2\dot{\varpi}_2 + l_3\dot{\varpi}_3 \approx 0,
\label{tbres}
\end{equation}
where $\lambda_{i}$ and $\varpi_{i}$ denote the mean and perihelion longitudes respectively. The indexes $(j_1,j_2,j_3) \in \mathbb{Z}^3\setminus\{0\}$ and $(l_1,l_2,l_3) \in \mathbb{Z}^3$ are conditioned by D'Alembert's rule:
\begin{equation}
\sum_{i = 1}^{3}(j_i + l_i) = 0.
\label{dalembert}
\end{equation}

For a specific three-body mean motion resonance (i.e. $\dot{\lambda}_i=n_i$ yields a  fixed value of $a_i$), eq. \eqref{tbres} defines several multiplets associated to different vectors of integers $\bm{l}$, each located at slightly different resonant values of the corresponding semi-major axis. These multiplets (or sub-resonances) will inevitably overlap, generating an extended chaotic region in the three-body resonance. The full $(j_1,j_2,j_3)$ MMR Hamiltonian $(P_1,P_2,P_3)$, up to second order in the eccentricity of the small body $(P_3)$ can be reduced to a four dimensional one. Indeed, after defining:

\begin{equation}
\bm{I}=\left(N_3, S_1,S_{2},S_{3}\right),\qquad\bm{\theta}=\left(\phi,\varpi_1,\varpi_{2},\varpi_{3}\right),  
\label{variables}
\end{equation}
where $\phi=j_1{\lambda}_1 + j_2{\lambda}_2 + j_3{\lambda}_3$ and $N_3=L_3/j_3$, then
following the approach of Section~\ref{chaos0}, the local Hamiltonian reads
\begin{equation}
\begin{split}
H(\bm{I},\bm{\theta})=&-\frac{1}{2j_3^{2}N_3^{2}}-\beta_{0}\left(1+\frac{S_3}{j_3N_3}\right)^{2} +\\ &\left(j_{1}n_{1}+j_{2}n_{2}\right)N_3 + 
\nu_{1}S_{1}+\nu_{2}S_{2}+V(\bm{I},\bm{\theta}),
\end{split}
\label{MMRH}
\end{equation}
where $\beta_0\sim e_{3}^2$, $\nu_{1,2}$ are perihelion motions of $P_1$ and $P_2$ massive planets respectively. The perturbation takes the form:

\begin{equation}
V\left(\bm{I},\bm{\theta}\right)=\sum_{\bm{l}}\beta_{\bm{l}}(\bm{I})
\cos(\phi+l_1\varpi_1+l_2\varpi_2+l_3\varpi_3),
\label{pertMMRH}
\end{equation}
and the small coefficients $\beta_{\bm{l}}(\bm{I})$ can be given in terms of a power series of the small body's eccentricity (see \citet{nesvorny_morbidelli_1999}).

\begin{figure}  
\centering
\includegraphics[scale=0.3]{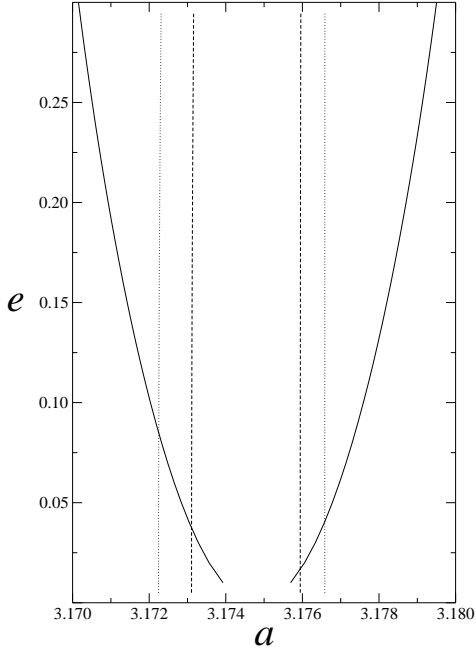}
\caption{Three resonances model for the $(5, -2, -2)$ three-body MMR. The strength of
each resonances is given by the corresponding width. The largest one corresponds to the resonance
$(5\lambda_J-2\lambda_S-2\lambda-\varpi)$ while the smallest one to $(5\lambda_J-2\lambda_S-2\lambda+\varpi_S)$.}
\label{multi}
\end{figure}

Considering three different multiplets of the asteroidal three-body resonance $(5, -2, -2)$, \citet{cachucho_etal_2010} applied Chirikov's diffusion theory to investigate, among other effects,  variations of the eccentricities of the (490) Veritas family. They clearly show that it is necessary to consider at least the three strongest terms in (\ref{pertMMRH}) in order to explain the observed distribution of eccentricities of this asteroidal family. This multiplet of three resonances for this particular MMR in the SS, given by $\bm{l}=(-1,0,0), (0,-1,0), (0,0,1)$, is represented in Fig.~\ref{multi}.

This simple model shows that the three resonances are in overlap and thus the full domain of the MMR is expected to be chaotic and therefore diffusion might occur. Moreover, the above figure
allows us to say that diffusion \emph{along} the resonance corresponds to variations of the
eccentricity while diffusion \emph{across} the resonance measures variations of the semi-major axis.

In the case of Gliese-876, $m_3\ll m_1 < m_2$, and since we are dealing with the full four dimensional resonant Hamiltonian in a small domain around Laplace resonance, a dense
set of resonances of the form 
$$\dot{\phi}_{lap}+l_1\dot{\varpi}_1+l_2\dot{\varpi_2}+l_3\dot{\varpi_3}\approx 0$$ 
would appear as well as many other nearby MMR. Thus regular motion is not expected in this region but a very complex domain of overlap of many resonances. Therefore the only way to investigate diffusion is by numerical experiments.

In order to understand the structure of the Laplace resonances and the role of the different resonances in the multiplet in the diffusion process, let
us write the Hamiltonian in Chirikov's style, taking again the same variables as defined
in (\ref{variables}) with $\phi=\phi_{lap}$.  Due to the D'Alembert's rule for the Laplace
resonance, the harmonic vector $\bm{m}\equiv(1,0,0,0)$ should be resonant, thus we take the
angle $\bm{m}\cdot\bm{\theta}=(\lambda_1-3\lambda_2+\lambda_3)$ as the resonant one. Taking
away from the perturbation the resonant term, the Hamiltonian becomes

\begin{equation}
\begin{split}
H_r(\bm{I},\bm{\theta})=&-\frac{1}{2j_3^{2}N_3^{2}}-\beta_{0}\left(1+\frac{S_3}{j_3N_3}\right)^{2} + \left(j_{1}n_{1}+j_{2}n_{2}\right)N_3 +\\ 
&\nu_{1}S_{1}+\nu_{2}S_{2}+\beta_{{\bm{m}}}(\bm{I})\cos(\bm{m}\cdot\bm{\theta})+V,
\end{split}
\label{H_r}
\end{equation}
where $V$ includes all terms of the form $\cos(\phi_{lap}+l_1\varpi_1+l_2\varpi_2+l_3\varpi_3)$ with $l_i\ne 0$. Following the formulation of Section~\ref{chaos1}, a canonical transformation or local change of basis $(\bm{I},\bm{\theta})\to(\bm{J},\bm{\psi})$ such that 
$$\psi_k=\sum_{i=1}^4 \mu_{ik}\theta_k,\qquad I_s=I^r_s+\sum_{k=1}^4J_k\mu_{ks},$$ 
where $\mu_{ik}$ are the coefficients of the transformation with $\mu_{1k}=m_k$, 
$\mu_{2k}=\omega^r_2/||\bm{\omega^r}||,\dots$, (so $\psi_1=\bm{m}\cdot\bm{\theta}$),
allows one to reduce the resonant Hamiltonian to
\begin{equation}
\begin{split}
 H(\bm{J},\bm{\psi})=&{J_1^2\over 2M} 
+|\bm{\omega}^r|J_2 + \sum_{s=1}^4 \sum_{k+ s> 2}^4 \frac{J_kJ_s}{2M_{ks}}+\\
&\beta_{{\bm{m}}}(\bm{I^r})\cos\psi_1+
\sum_{\bm{l}} \beta_{\bm{l}}(\bm{I^r})
\cos(\bm{l}\cdot\bm{\theta}(\bm{\psi})),
\end{split}
\label{Hfull}
\end{equation}
where $M$ is the non-linear mass defined in Section~\ref{chaos1} while the
$M_{ks}$ are similar constants to $M$ but involving different coefficients of the basis
transformation $(\mu_{ik})$ and
$\bm{I^r}$ is the resonant action that satisfies the resonance condition
$$n_1(\bm{I^r})-3n_2(\bm{I^r})+2n_3(\bm{I^r})=0.$$ 

Recalling that the dot product is invariant, the replacement $\bm{\theta}\to\bm{\psi}$ is easily done since $\bm{l}\cdot\bm{\theta}=\bm{r}\cdot\bm{\psi}$, where now the components of $\bm{r}$ are real numbers.

Keeping only the actual resonance ($\phi_{lap}$) and neglecting all the perturbation terms $\beta_{\bm{l}}(\bm{I^r})$  for  all $\bm{l}$,  the components $J_2,J_3, J_4$ become local integrals of motion whose value is equal to zero if $\bm{I^r}$ is a point of the orbit. Then, the Hamiltonian reduces to a pendulum-like model

\begin{equation}
\tilde{H}_r(J_1,\psi_1)=\frac{J_1^2}{2M}+V(\bm{I^r})\cos\psi_1.
\label{pen2}
\end{equation}

Thus the motion \emph{across} the resonance is given by $J_1$, the pendulum action.
It librates or circulates depending on the value of $\tilde{H}_r$ and for $\tilde{H}_r=V(\bm{I^r})$ the system lies at the separatrix.  When switching on the perturbation $(\beta_{\bm{l}}(\bm{I^r})\ne 0)$ the main effect to the pendulum model is to produce a distortion of the separatrix
and the motion in the neighborhood of this asymptotic trajectory becomes chaotic leading to
the so-called chaotic layer. However a non-vanishing perturbation, including at least two perturbing terms, also leads to variation
of the unperturbed local integrals $J_2,J_3,J_4$, after a simple inspection of (\ref{Hfull}).
The variation of $J_2$ has a direction normal to the energy surface and thus it can be ignored.
Changes in $J_3$ and $J_4$ lie in the diffusion space and therefore \emph{along} the resonance.
In other words,  due to the particular geometry of the resonance (see Fig.~\ref{multi}), $J_1$ measures diffusion in the  semi-major axis of $P_3$ while, $J_3$ and $J_4$ lying in the diffusion space, take into account  diffusion in the eccentricity of the small body.

From the above discussion it becomes clear that if $\bm{m}\cdot\bm{\theta}=(\lambda_1-3\lambda_2+\lambda_3)$ is a resonant angle, then $(\lambda_1-3\lambda_2+\lambda_3+l_1\nu_1+l_2\nu_2+l_3\nu_3)$ is also resonant for any integers $l_i\ne 0$ that satisfy the D'Alembert's rule. And as we have already shown, all these resonances are overlapping since all of them have almost the same $\bm{I^r}$ (or $a^r$). Hence we expect a fully chaotic domain within the Laplace resonance and therefore diffusion in both directions, along and across the resonance. Moreover, since many other MMR are very close to this Laplace resonance, a large chaotic sea should surround it. As can be seen in Figs.~\ref{fig2}, the correspondence between the simple model and the full numerical experimentations is, at least qualitatively, most evident. All this is what we observe in Figs.~\ref{fig2}. However, from the above discussion, nothing could be said about the diffusion rate or if the diffusion has a normal character.

\begin{figure*}
\centering
\mbox{\includegraphics*[width=16.0cm]{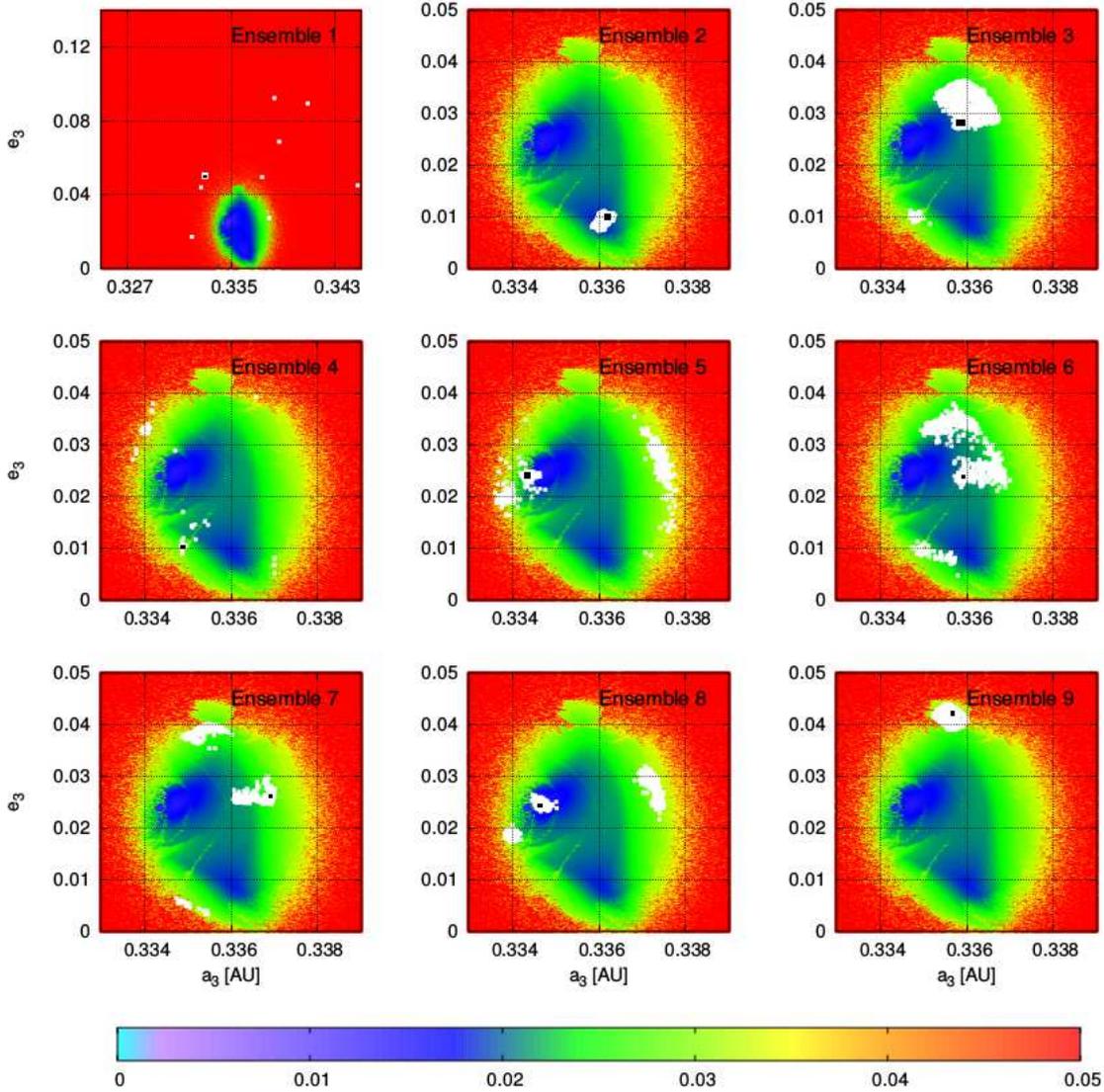}}
\caption{Diffusion of 9 ensembles of 256 initial conditions defined in different regions of the representative plane. Total integration time was $2 \times 10^5$ yrs. Black rectangles show the location of the initial ensembles, while the color dots indicate their diffusion during this time-span.}
\label{fig3}
\end{figure*}

\section{Diffusion Inside the Laplace Resonance}

Having analyzed the general structure and chaoticity of the Laplace resonance, our next step is to estimate the diffusion times in the different regions within this commensurability.

We performed a series of integrations of ensembles of initial conditions at specific locations in the $(a_{3},e_{3})$ plane. Each ensemble consisted of a total of 256 initial conditions, all centered around a given point in the plane, and defining very narrow regions of at most $10^{-3}$ in $\Delta e_3$ and $2\times 10^{-4}$ in $\Delta a_3$. Each initial condition was again integrated for a total time of $2\times 10^{5}$ years, twice longer than the time-span used for the original map. 

During the evolution, we kept a record of every time the particles crossed the representative plane. This was said to occur when the following conditions were satisfied:

\begin{itemize}
\item $\Sigma_{i=1}^{3} (|M_{i}-M^{0}_{i}| + |\varpi_{i}-\varpi^{0}_{i}|) < \epsilon_{ang}$ ,

\item $\Sigma_{i=1}^{2} |e_{i} - e^{0}_{i}| < \epsilon_{e}$ ,

\item $\Sigma_{i=1}^{2} |a_{i} - a^{0}_{i}| < \epsilon_{a}$,
\end{itemize}
where $\epsilon_{ang}$, $\epsilon_{e}$ and $\epsilon_{a}$ are predefined values. For this set of simulations we adopted $\epsilon_{ang} = 6^{\circ}$, $\epsilon_{a} = 0.005 \textrm{AU}$ and $\epsilon_{e} = 0.005$ .

We integrated a set of nine ensembles (hereafter referred to as 1S, 2S, ... , 9S). The first was located in the outer resonant region, while the other ensembles were placed inside the inner resonant region.

Figure \ref{fig3} shows the evolution of each ensemble in the representative plane, superimposed with the resonant structure as determined by the $\Delta e_3$ indicator. The initial conditions are indicated with black rectangles while their subsequent diffusive evolution is depicted in colors. 

As seen from the left-hand plot, the S1 ensemble suffers a large-scale diffusion, rapidly covering all the outer resonant region. Motion is highly chaotic and the times between crossings are unpredictable. More interesting, all intersections with the representative plane occur in the red region, which appears detached from the inner resonant zone indicated in green and blue. 

The remaining frames in Figure \ref{fig3} correspond to initial conditions in the inner resonant zone. In all cases the diffusion is very localized, at least compared with the evolution of 1S. Moreover, the time evolution of the ensembles never leave the inner domain, indicating no noticeable mixing between both parts. This seems to suggest that perhaps both regions are dynamically unconnected (at least up to the considered length of the simulations) and that the limit between them could represent a kind of dynamical boundary inside the resonance. In consequence, initial conditions within the inner region seem to be characterized by very small diffusion rates. The opposite seems to occur for initial conditions in the outer domain.

\subsection{Diffusion Coefficients}

In this Section, we proceed to quantify the different chaotic regimes within the Laplace resonance. 
To this end we take advantage of the ensembles 1S to 9S described in the previous section to ensure a sufficiently representative ensemble to compute the variances in both $a_{3}$ and $e_{3}$. The ensembles labeled as $i = 2, ... , 9$ have a considerable number of intersections with the $(a_3, e_3)$ plane that ensure a sufficiently good approximation to the actual experimental value of the variance of $(a_3,e_3)$. In the case of S1, we already noticed the difficulty of having a significant amount of crossings with the representative plane.

The numerical computation of the variance proceeds as follows: {\bf i)} We subdivided the total integration time of the ensembles (i.e. $T_{tot} = 2\times 10^5$ years) into $N_t$ time intervals of fixed length $T_{imp}$, so that $T_{imp} = T_{tot}/N_t$. {\bf ii)} At each time interval $[(i-1) T_{imp},i T_{imp}], \quad i = 1, N_t$ we computed the total plane crossings conditions $(N_{i})$ which occur at shorter times than the extreme value of the time interval (i.e. if $T_{cr} < i T_{imp}$). {\bf iii)} A representative value of the variances both, for $a_{3}$ and $e_{3}$ is calculated using all the plane crossing conditions in each time interval following:

\begin{equation}
\sigma_x = \frac{1}{N_i}\Sigma (x(T_{cr}) - x_{0})^2,
\end{equation}
were $x$ should be replaced by any of the fundamental parameters $a_{3}$ or $e_{3}$, and $x_{0}$ is either the semi-major axis $a_{3}$ or the eccentricity $e_{3}$, at the center of the ensemble.

Diffusion processes are commonly characterized by a power law relationship of the form $\sigma^2(t)=c~t^\alpha$, with $c>0$. If $\alpha=1$ we have normal diffusion, while in case of $\alpha<1$ the phenomenon is called  sub-diffusion, or when $\alpha>1$ it is called super-diffusion. In the normal diffusion case, that corresponds to purely random motion, it is possible to define a numerical diffusion coefficient, $D$, as the constant rate at which the variance grows with time. The computation of the diffusion coefficient in case of sub-diffusion or super-diffusion for a generic HS is yet an open and difficult
problem.  Therefore in this work we focus on which type of diffusion dominates the different regions of the resonance discussed above. 

Thus we associate to $\sigma_x^2$ a power law
\begin{equation}
  \label{intro_dif_coef_variance}
  \sigma_x^2(t) =  c_x t^{\alpha_x},
\end{equation}

where $c_x$ and $\alpha_x$ are the fitted parameters. In case of an exponent $\alpha\approx 1$ the parameter $c_x$ is an estimate of the actual and standard diffusion rate coefficient, $D_x$ . On the other hand, if $\alpha$ is far from $1$, nothing could be said about the diffusion coefficient. 
Only a qualitative description about the diffusion processes in phase-space could be provided.

\begin{figure}
\centering
\mbox{\includegraphics[width=8.0cm]{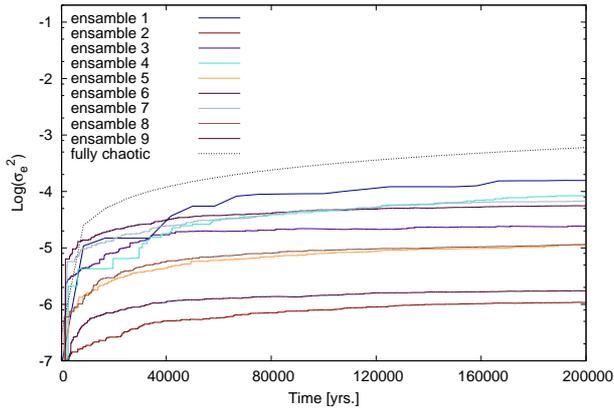}}
\caption{Variance of the eccentricity as function of time, obtained for each of the ensembles 1S to 9S. The $\sigma_e$ values are shown in logarithmic scale in order to see how each curve moves away from the Normal Diffusion curve, depicted in dashed black line on the plot.}
\label{fig4}
\end{figure}

In Figure \ref{fig4} we show the time-evolutions of $\sigma_e$ for each of the ensembles 1S - 9S. We also include in the figure the corresponding time-evolution for the completely random case ($\sigma^2\propto t$) just for the sake of comparison. The figure shows that in all cases the nine ensembles have a smaller rate than the expected one for normal diffusion. The ensemble 1S, taken in the outer part of the resonance, presents some similarities with the normal case.  For the rest of the ensembles, the 4S shows the highest rate of evolution at large times, but the computed variance for this ensemble is one order of magnitude less than that for the ensemble 1S. This result clearly shows that the inner region of the resonance, while chaotic, presents a dynamical behavior that looks almost stable and therefore the diffusion, is not well approximated as a Brownian type motion. In Table~\ref{table3} we show the values of these exponents for the nine ensembles. Clearly only the ensembles 1S and 4S present an exponent close to 1. The associated diffusion coefficient, obtained by a linear fit for $t\gtrsim 10^4$ years results $D\sim 10^{-9}$ for both ensembles. The rest of the ensembles are highly sub-diffusive and thus the dynamical behavior is rather stable at least for $t=2\times 10^5$ years. The particular case of ensemble 4S might be explained since its initial position $(a_0, e_0)\approx (0.335, 0.01)$ is also in the outer region of the resonance but very close to the boundary defined by the MEGNO computation (see for instance Fig.\ref{fig5}). Initially the evolution of the ensemble shows a sub-diffusive behavior but, for moderate times, the diffusion becomes more normal and maybe for larger times it could reach higher values of the eccentricity.

In this direction, \citet{batygin_holman_2015} developed a 2-Dimensional model and studied the diffusion on this same system. Besides the difference between the analytic and numerical approximations, they assume a normal diffusion to derive a diffusion coefficient. Our approach shows that this assumption for diffusion in a multi-resonant system such as GJ-876 is not well suited, at least in the inner region of the Laplace resonance.

\begin{table}
\centering
\begin{tabular}{ c c }
\\[1ex] 
\hline\hline \\[-1.3ex]
{\bf Ensemble} & $\alpha$\\
\hline \\
   1S  & $0.942715$ \\ 
   2S  & $0.585784$ \\
   3S  & $0.494802$ \\ 
   4S  & $0.923109$ \\ 
   5S  & $0.648737$ \\
   6S  & $0.448689$ \\
   7S  & $0.686534$ \\
   8S  & $0.592316$ \\
   9S  & $0.462431$ \\
\hline
\end{tabular}\label{table3}
\\[1ex]
\caption{
Exponents $\alpha$ calculated by a least-squares fit for the data obtained by the variances from each of the nine ensembles.}
\end{table}

\section{Orbital Stability in the Inner and Outer Resonant Regions}

\begin{figure}  
\centering
\mbox{\includegraphics*[width=9.0cm]{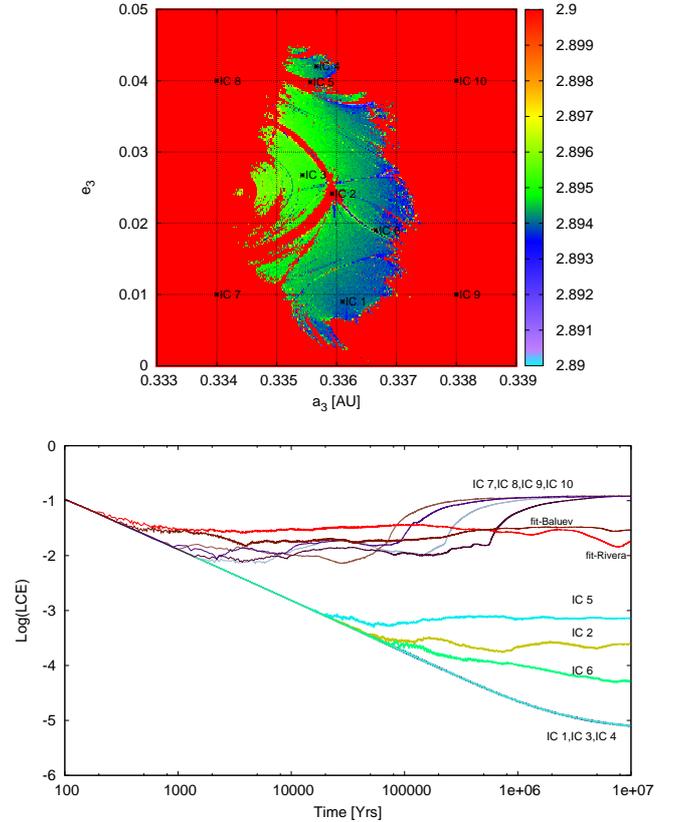}}
\caption{Bottom frame shows the Maximum Lyapunov Coefficient (LCE) for 10 initial conditions chosen in different regions of the representative plane (identified in top graph).}
\label{fig5}
\end{figure}

Finally, we wanted to analyze the orbital stability and dynamics of a set of initial conditions in both regions of the Laplace resonance. Figure \ref{fig5} shows the Lyapunov characteristic exponent (LCE) calculated for 10 initial conditions in the representative plane. Their locations, supper-imposed to the Megno-map, are shown in the upper frame, while the time evolution of their LCE is shown in the bottom plot. Along with the evolution of their LCE, we have plotted the corresponding evolution of the initial conditions represented by the co-planar orbital fits already mentioned in section \ref{sec4.1}.

\begin{figure}  
\centering
\mbox{\includegraphics*[width=9.0cm]{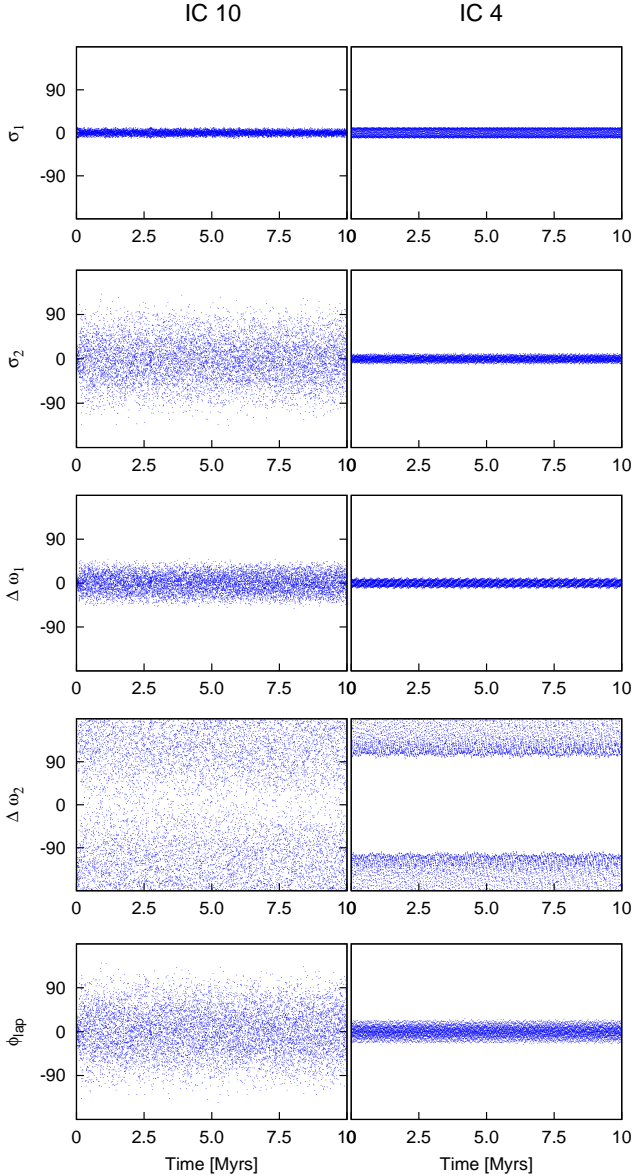}}
\caption{Time evolution of the resonant angles corresponding to initial conditions IC 10 and IC 4 described in Figure \ref{fig5}. These show the evolution two characteristic conditions placed at the inner (IC 4) and at the outer (IC 10) resonant regions.}
\label{fig6}
\end{figure}

All the initial conditions set in the outer resonant region (IC 7 through IC 10) are characterized by very large values of LCE, of the order of $10^{-1}$ yrs$^{-1}$, corresponding to extremely chaotic motion. However, as shown in the left panels of Figure \ref{fig6} for IC 10, there is no indication of orbital instability, at least within several $10^7$ years. The system is inside the Laplace resonance, although the resonant angle displays large-amplitude librations. The resonant angles of the individual two-body resonances are also librating, and the behavior of $\Delta \varpi_1$ indicates that $m_1$ and $m_2$ are trapped in an Apsidal Corotation Resonance (ACR) \citep{beauge_etal_2003}. The difference in longitudes of pericenter of the outer pair ($\Delta \varpi_2$), however, circulates, indicating that this sub-system is not in an ACR. 

These values of LCE are very similar to those obtained by \citet{batygin_holman_2015}, where they estimate a lyapunov time for Rivera's orbital fit using the aforementioned 2-dimensional model. In fact, all orbital fits show a similar behavior (see Figure \ref{fig5}), with values of LCE somewhere between those corresponding to the IC7-IC10 and IC2-IC5-IC6 groups of initial configurations.

Continuing with Figure \ref{fig5}, initial conditions placed in the red streaks within the inner resonant region (IC 2, 5 and 6) have moderate values of LCEs. While these are significantly smaller than before, they still correspond to significant chaotic motion. Finally, the initial conditions placed inside the relatively regular inner resonant region (IC 1, 3 and 4) all show almost identical very small values of the Lyapunov exponent. At the end of the simulation, at $T=1.2 \times 10^7$ years, the value of LCE has yet to reach a plateau, indicating that this region is characterized by very regular motion. Indeed, the theoretical expected final value of the LCE for regular motion is $\ln T/T\sim 10^{-6}$.

The right-hand frames of Figure \ref{fig6} shows the time evolution of the resonant angles for IC 4. All, together with differences in longitudes of pericenter, exhibit small-amplitude librations, indicating that this configuration is not only trapped inside the Laplace resonance but also exhibits a double-ACR. The same is noted for the other initial conditions in this region. This seems to indicate that the difference in dynamics between the inner and outer resonant domains is defined by the behavior of the auxiliary resonant angles, particularly that of the outer pair. Thus, it appears that the almost regular region deep within the Laplace resonance corresponds to Double-ACR orbits, while the highly chaotic outer region is associated to an ACR for the inner pair and a $\sigma_2$-libration of the outer pair of planets.

\section{Conclusions}

The choice of the GJ-876 system, although arbitrary, is due to two main factors. On one hand, we want to analyze the diffusive process and chaotic mixing in a system which could have avoided other chaotic processes during its early stages after gas depletion. In this sense, GJ-876 is a well characterized system that displays a resonant chain of planetary bodies. On the other hand, a natural motivation was the extensive quantity of previous works where this specific planetary system has been used as prime example. 

We have started our analysis by improving our representation of the region covered by the Laplace resonance in the $(a_{3},e_{3})$ plane. We integrated one order of magnitude more initial conditions than we had previously done, and also extended the total integration time for each to 2$\times 10^{5}$ years. We therefore explored in a very precise way the main dynamical structures that this system represents. 

As was already pointed out in \citet{marti_giuppone_beauge_2013}, we recognized two main regions in the surroundings of the resonance. The one we called inner resonant region is characterized by lower values of $\Delta e_{3}$, a MEGNO indicator value of $\langle Y \rangle \sim 2$ and utterly very small values for the LCE which result in seemingly large lyapunov times. The outer resonant region is, however, dominated by extremely chaotic dynamics, presenting high values of $\Delta e_{3}$ and $\langle Y \rangle$, and having LCE's somewhat higher than in the inner region. Moreover, we also concluded that the inner zone corresponds very well with the region of lower libration amplitude of the resonant angle $\phi_{lap}$. This feature, although trivial, is extremely important because it shows that the multi-resonant configuration of the four-body system ($m_{0} + m_{i}, i = 1, 3$) is responsible for its long-term stability. The coincidence in the low-amplitude libration regions of $\sigma_{2}$ and $\phi_{lap}$ on the phase-space allows us to state that the system is unable to show a libration of the Laplace angle without being trapped in the two single two-body resonances.

Although both the MEGNO and $\Delta e_{3}$ indicators point towards chaoticity within the inner resonant region, this characteristic should be considered with care. Indeed, we have already stated that aside the overall chaoticity of the system, we could still define regions with completely different dynamical behaviors. The higher precision used in our grid-simulations of initial conditions allowed us to perform a much more detailed map of the inherent chaotic structure inside the Laplace resonance. As it was shown in the bottom frame of Figure \ref{fig2}, several thin strips of higher values of $\langle Y \rangle$ cross each other along the whole inner domain. This behavior is completely expected (see section \ref{sec4.2}) due to  the overlapping of resonances associated with slight variations of the the longitudes of perihelion of the planets, which are located at the same region as the Laplace resonance.

In order to get a quantitative idea of how the different aspects of chaotic behavior affect the dynamics of the system and its resonant structure, we performed numerical calculations concerning the diffusive process, which occur inside the multi-resonance domain. Although diffusion is always present, we show here that the rate at which the local variation of fundamental parameters $(a_{3},e_{3})$  associated with the actions in phase-space (see section \ref{model}), is completely limited to the inner region of the resonance as long as their initial values reside in that domain. In a few cases where the initial conditions were located at the borders of the inner region or at the strips of moderate chaos, the diffusion rate seems to be higher. We also performed calculations of the diffusive process for an ensemble of initial conditions located outside the inner resonant region, yielding a time-evolution of the variance very close to normal diffusion (i.e. $\alpha = 0.942715$ in the model $\sigma^2(t)=c~t^\alpha$), while  for any of the other ensembles the fit of this exponent was seemingly smaller. This result clearly shows that the assumption of normal diffusion $(\sigma_2(t)\propto t$) for these kinds of systems is not well sustained.

The LCE calculated for 10 different initial conditions, chosen to represent some crucial aspects of the resonance, are clearly in accordance with the overall analysis developed here. There is a direct link between the lower values of LCE and initial conditions at the inner zone. Accordingly, for systems with initial conditions placed outside the inner part, they not only reached higher values of LCE, but they also reach these values at earlier times than systems with initial conditions at the inner region. Moreover, Those conditions which were located specifically at the moderate MEGNO strips, show an intermediate value of LCE, and even some, seem not to have reached its asymptotic LCE value at the final time of the simulation. 

The LCE obtained by \citet{batygin_holman_2015} corresponds to the outer resonant region of the Laplace resonance, as they make use of the fit from \citet{rivera_etal_2010} (see also table \ref{table2}). However, we found that the inner region, characterized by a Double-ACR and small amplitude of librations of the resonant angles, contains initial conditions which are less chaotic, associated with Lyapunov times larger than $10^5$ years. In fact we have run a simulation of Rivera's orbital fit, which led to a Lyapunov time of $\sim 100$ years. Our integrations for initial conditions in the inner resonant region which are not specifically on any of the moderate MEGNO strips, are not only stable for more than $10^{7}$ years, they also show a much more limited evolution of the libration amplitudes of the resonant angles (see right-hand frame of Figure \ref{fig6}), as well as a much regular variation. This strongly suggests that although chaotic, the system could and in fact has long-term stability, and that chaotic mixing should not have occurred in systems which display resonant dynamics similar to that of GJ-876.

Although this research was developed for a specific planetary system, it seems reasonable that the main characteristics of any system representing similar multi-resonant configurations could share the main features that were described throughout this paper. The implementations, although numerically expensive, should not carry major problems, and so, an extension to any such a system would only need a sufficiently precise orbital fit. As the number of multi-resonant systems is constantly increasing, this type of dynamical study is of fundamental importance mainly for stability considerations, and secondly because of the constrains that multi-resonant planetary systems can impose on the planetary formation theories.

\section*{Acknowledgments}

This work used computational resources from CCAD – Universidad Nacional de C\'ordoba (http://ccad.unc.edu.ar/), in particular the Mendieta Cluster, which is part of SNCAD – MinCyT, Rep\'ublica Argentina. Other Numerical simulations were made on the local computing resources from the Instituto de Astronom\'ia Te\'orica y Experimental (IATE), at the University of C\'ordoba (C\'ordoba, Argentina) and also on the IFLySIB computational resources at the Instituto de F\'isica de L\'iquidos y Sistemas Biol\'ogicos - CONICET - UNLP. We also want to thank the Instituto de Astrofísica de La Plata (IALP) - CONICET - UNLP, La Plata, for supporting this research.

The authors also wish to express their gratitude to an anonymous referee for important suggestions and comments.

\bibliographystyle{mnras.bst}

\bibliography{library}

\end{document}